\documentclass[aps,pra,twocolumn,groupedaddress,nofootinbib]{revtex4-2}
\usepackage{amsthm}
\usepackage{graphicx}
\begin{document}

%\preprint{}

%\title{Locality claims based on Bell inequality inconsistency are inconsistent}
\title{Comment on ``Nonlocality claims are inconsistent with Hilbert-space quantum mechanics''}

\author{Justo Pastor Lambare}
%\email[]{jupalam@gmail.com}
%\thanks{}
\affiliation{Facultad de Ciencias Exactas y Naturales, Ruta Mcal. J. F. Estigarribia, Km 11 Campus de la UNA, San Lorenzo-Paraguay}


\date{\today}

\begin{abstract}
The view exists that the Bell inequality is a mere inconsistent application of classical concepts to a well-established quantum world.
In the article, ``Nonlocality claims are inconsistent with Hilbert-space quantum mechanics'' [Phys. Rev. A, 101, 022117, (2020)], Robert B. Griffiths advocates for quantum theory's locality.
Although R. B. Griffiths presents valuable insights in favor of quantum mechanics' local character, he argues the Bell inequality is an inconsistent application of classical physics to quantum mechanics.
We explain why a correct assessment of the  Bell inequality does not in fact conflict with Griffiths's views of quantum locality and, on the contrary, it already contemplates them.
Hence, Bell inequality inconsistency is not necessary for Griffiths's quantum locality to hold.
\end{abstract}

%\keywords{}

\maketitle
\section{Introduction}\label{sec:INTRO}
The Bell inequality transcends the field of purely philosophical interest and quantum foundations.
Far from being an obsolete subject, it is an active component of present and future quantum information technology. Whence its correct assessment is indeed relevant.

In Ref. \cite{pGri20}, Robert Griffiths advocates for the local character of quantum theory.
He bases his arguments on two concepts: his measurement framework (consistent histories) and the no-signaling principle.

Those arguments can be accepted or rejected but are not themselves incorrect.
These kinds of apparent ambiguities are common when addressing a highly controversial subject like quantum mechanics interpretation.

Griffiths's views of quantum locality reject the very hypotheses that underlie the Bell theorem.
For that reason, the inequality violations do not pose any threat to Griffiths's views of locality.

We purport to show how Griffiths's measurement framework and the no-signaling principle coherently fit into Bell's perspective of nonlocality.

The Bell theorem's hypotheses concern physical concepts, like locality and freedom, equally applicable to classical physics and quantum mechanics.
The concept of locality relevant to the discussion of the Bell inequality is what Bell called local causality \cite{pBSHC85}.
We shall refer to this locality concept as \emph{Bell-Local Causality} (BLC).

On the other hand, we shall see that Griffiths upholds two different notions of locality: one is a \emph{No-Signaling locality} (NSL), and the other is coincident with BLC.
As we anticipated, although both locality concepts are different, they involve no inconsistencies between quantum mechanics and Bell's formulation.

First, we review the meaning of the Bell inequality then go over Griffiths's arguments.
We make clear that we do not advocate for quantum nonlocality, neither we consider Griffiths's arguments of quantum locality as erroneous.
Our intention is to stress why it is misleading to consider the Bell inequality as an inconsistent application of classical physics to Hilbert-space quantum mechanics.
\section{Meaning of the Bell inequality}\label{sec:MOBI}
John Bell first conceived his theorem in 1964 \cite{pBel64} as a continuation of the \emph{Einstein-Podolski-Rosen} (EPR) \cite{pEPR35} argument.
Bell investigated the possibility to locally explain the existence of perfect correlations predicted by quantum mechanics.
We shall consider the \emph{Clauser, Horne, Shimony, Holt} (CHSH) \cite{pCHSH69} generalization of the Bell inequality.

In 1975 \cite{pBSHC85}, Bell gave a formulation that rests on two hypotheses; BLC and freedom.\footnote{Notice the absence of metaphysical assumptions such as counterfactual definiteness, elements of physical reality, incompatible experiments, etc. Although Bell never used them, they are more obviously irrelevant in his 1975 formulation. All these metaphysical extravaganzas are related to the unnecessary and inconsistent use of counterfactual reasoning, see Refs. \cite{pLam17,pLam21a,pLam21b,pLam20b}.}
Hence, violation of the inequality implies the violation of at least one of those assumptions.

BLC and freedom are either true or false.
They apply to natural phenomena irrespective of the theory explaining them.
They cannot hold consistently true in a quantum sense and at the same time be classically false or vice versa.

J. Jarret \cite{bShi84b} proved that BLC is the conjunction of two conditions usually known as parameter independence and outcome independence.
Freedom usually appears under the measurement independence or statistical independence hypothesis.
Whether statistical independence is indeed necessary for freedom is another debate that does not concern us here \cite{bHal16,pHos20}.

For our purposes, we do not need to elaborate more on those concepts. The interested reader can find detailed explanations in the literature, for instance, Ref. \cite{bHal16}.
\section{Noncommuting Operators}\label{sec:NCO}
Here we analyze the claim that violations of the Bell inequality only prove the nonexistence of commuting operators and have nothing to do with nonlocal effects on distant particles.

To prove his thesis, Griffiths conceived an experiment with neon atoms violating a Bell-type inequality, where locality is not an issue.

In Griffiths's ingenious experiment, only one particle produces a Bell-type inequality, so obviously, locality is not an issue.
Although locality is not an issue when measuring only one particle, it certainly is when simultaneously measuring two different particles far apart.
Thus, if Griffiths's point is to prove the Bell inequality emerges under different circumstances having interpretations that are not related to locality, his example is, indeed, correct.
However, that does not prove locality cannot be involved in other contexts where the inequality arises, for instance, a CHSH spin correlation experiment.

Interestingly, other authors have proposed similar ideas for interpreting the inequality violations trying to escape from the grip of the Bell inequality.
For instance, Andrei Khrennikov \cite{pKhr19} proposes the same interpretation as Griffiths's but applied in a CHSH context.
According to A. Khrennikov, \emph{``We demonstrate that the tests on violation of the Bell-type inequalities are simply statistical tests of local incompatibility of observables''}.
Khrennikov shows that the Bell operator
\begin{equation}\label{eq:BO}
\mathcal{B}= A_0B_0+A_0B_1+A_1B_0-A_1B_1
\end{equation}
satisfies
\begin{equation}\label{eq:Krel}
\mathcal{B}^2=4I-[A_0,A_1][B_0,B_1]
\end{equation}
(\ref{eq:Krel}) means that when at least one commutator on the RHS vanishes, we obtain the Bell inequality.
Let $CO$ stand for the existence of at least one pair of commuting operators, then according to (\ref{eq:Krel}), Khrennikov proved that
\begin{eqnarray}
CO &\rightarrow BI\label{leq:cobi}
\end{eqnarray}
If $SI$ stands for statistical independence, Bell proved that
\begin{eqnarray}
BLC\wedge SI &\rightarrow BI\label{leq:lrbi}
\end{eqnarray}
Although in Griffiths's example, (\ref{leq:lrbi}) does not make sense, in a CHSH experiment, where (\ref{leq:cobi}) and (\ref{leq:lrbi}) are both applicable, Khrennikov suggests that by proving (\ref{leq:cobi}), he has disproved (\ref{leq:lrbi}).
Such arguments ignore the logical fact that a statement may have different, sometimes unrelated, sufficient conditions.
None of them disproves the validity of the others as a sufficient condition.
Of course, when analyzing (\ref{leq:cobi}), nonlocality is not an issue.
Likewise, when analyzing (\ref{leq:lrbi}), operators' commutativity is not an issue.

According to (\ref{leq:cobi}), a Bell inequality violation means that $[A_0, A_1]\neq0$ and $[B_0, B_1]\neq0$.
At the same time, according to (\ref{leq:lrbi}), either BLC or SI is false .

No matter what other inequality interpretations we may find, they do not invalidate the locality implications of a CHSH singlet state correlation experiment \cite{pSza94,pLam21b}.

For instance, some have pointed out that George Boole first discovered the Bell inequality in the mid-eighteen hundreds \cite{pKup18}.
Boole showed that a Bell-type inequality is a necessary condition for the existence of a joint probability(JP).
\begin{equation}\label{leq:jpbi}
JP\rightarrow BI
\end{equation}
Then again, the argument goes, violations of the Bell inequality are nothing else than proofs of the nonexistence of a joint distribution for the experiment's probabilities.

Whereas the previous interpretation is correct, the problem resides in the ``nothing else'' clause.
They also prove the nonexistence of commuting operators and invalidates the conjunction of Bell local causality and statistical independence when these concepts are applicable.

In the past, many researchers have observed the possibility of deriving the Bell inequality without assuming locality, hidden variables, or causality and then erroneously jumped to conclude the Bell inequality irrelevance regarding those matters \cite{pSza94}.
\section{Classical Hidden Variables}\label{sec:CHV}
Equation twenty-four of Griffiths's paper reproduces the factorization condition necessary to derive the Bell inequality
\begin{equation}\label{eq:Gri24}
Pr(A,B|a,b)=\sum_\lambda Pr(A|a,\lambda)Pr(B|b,\lambda)Pr(\lambda)
\end{equation}
Griffiths presents a detailed explanation of what is wrong with (\ref{eq:Gri24}).
He summarizes the argument in the final sentence of the corresponding section of his paper:
\begin{quote}
\emph{``Thus the usual derivations of CHSH and other Bell inequalities employ classical physics to discuss quantum systems, so it is not surprising when these inequalities fail to agree with quantum predictions, or the experiments that confirm these predictions.''}
\end{quote}

Griffiths, and usually quantum localists \cite{pWer14,pBou17}, consider the presence of the $\lambda$ common causes variables in (\ref{eq:Gri24}) as a hallmark of classical physics.
Although many non-localists \cite{pNor07,pGis12,pMau14a,pLau18} seem to contend such interpretation, we agree with Griffiths, and in general with quantum localists, that the inclusion of additional parameters foreign to quantum mechanics, in (\ref{eq:Gri24}), is a sign of ``classicality''.

However, discussing the classical nature of (\ref{eq:Gri24}) avoids the real problem that gave it origin.
Really, in 1975 Bell \cite{pBSHC85} gave a modified and improved version of the original 1964 \cite{pBel64} presentation of his theorem.
In his new formulation, Bell defined the BLC concept and explained that quantum mechanics is nonlocal because it violates BLC, not because it violates his inequality.

Indeed BLC is based on \emph{Reichenbach Common Cause Principle} \cite{pCav14}, i.e., locality requires correlations to be explained by a complete determination of the common causal past ``$\lambda$'' of the measuring events
\begin{eqnarray}
Pr(A,B|a,b,\lambda)  &=& Pr(A|a,\lambda) Pr(B|b,\lambda)\label{eq:BLC}
\end{eqnarray}
Testing quantum mechanics against BLC for the singlet state correlations only requires putting $\lambda=|\psi\rangle$ in (\ref{eq:BLC}), where $|\psi\rangle$ represents the ``quantum'' singlet state.
Then setting $A=1, B=1,a=b$, quantum mechanics predicts
\begin{eqnarray}
Pr(1,1|a,a,|\psi\rangle)  &=& Pr(1|a,\lambda) Pr(1|a,|\psi\rangle)\label{eq:BLC1}\\
0                    &=& \frac{1}{2} \frac{1}{2} =\frac{1}{4}\label{eq:BLC2}
\end{eqnarray}
Since $0\neq 1/4$, and we did not introduce additional variables foreign to quantum theory or invoked any particular interpretation of the wave function, then as Bell said \cite{pBSHC85}:
\begin{quote}
\emph{Ordinary quantum mechanics, even quantum field theory, is not locality causal in the sense of (2).}
\end{quote}
(2) refers in Bell's paper to BLC.\footnote{Bell did not use equations. He contented with a rather qualitative argument similar to the one Einstein gave in 1927 \cite{bBac09}.}
Thus, when quantum mechanics is considered complete, it becomes notoriously nonlocal in the BLC sense.

Notice that Bell did not even mention his inequality to prove that ordinary quantum mechanics is not locally causal.
Then the question arises, what's the use of the Bell inequality?
Well, to be sure, the Bell inequality does not prove that quantum mechanics is not locally causal.
It only proves that it is not possible to introduce common causes, satisfying the statistical independence assumption, to find locally causal (in the BLC sense) explanations of quantum mechanics strong correlations.

Claiming that quantum mechanics appears non-locally causal only because (\ref{eq:Gri24}) is a classical equation is untenable.
The last point is an endemic inconsistency affecting quantum localists' arguments who pretend to reject only the ``realism'' part of the compound expression local realism.\footnote{Although realism is an obscure concept \cite{pNor07}, there is one way in which its rejection makes sense. It is when it means the presence of a causal mechanism. Accepting that strong correlations do not need causal explanations, there would be no need for the existence of superluminal interactions \cite{pFra82}. Notice that this rational meaning of realism has nothing to do with metaphysical concepts such as elements of physical reality, counterfactual definiteness, etc. It is incorrect to ascribe such blunders to John Bell.}
The issue generated longstanding debates \cite{pNor07,pGis12,pWer14,pMau14a,pLau18,pBou17}.

Respecting Griffiths's views, there is a coherent way to reject the implications of (\ref{eq:BLC}), and it is not the classical character of (\ref{eq:Gri24}).
Griffiths clearly rejects no-signaling effects as being nonlocal influences:
\begin{quote}
\emph{``To be sure, those who claim that instantaneous nonlocal influences are present in the quantum world will generally admit that they cannot be used to transmit information; this is known as the ``no-signaling'' principle, widely assumed in quantum information theory.
This means that such influences (including wave-function collapse) cannot be directly detected in any experiment. The simplest explanation for their lack of influence is that such influence do not exist.''}
\end{quote}
That constitutes a rejection of Bell's local causality.
BLC given by (\ref{eq:BLC}) is the conjunction of two conditions: parameter independence and outcome independence \cite{bShi84b}.
Griffiths rejection of no-signaling effects as nonlocal influences implies the rejection of outcome independence and the acceptance of uncontrollable nonlocality.
Hence, (\ref{eq:BLC}) is not applicable, and quantum theory becomes a local theory according to NSL.

In conformity with NSL, (\ref{eq:Gri24}) does not apply, and its classical character is not an issue.
The issue at stake is whether we accept or reject outcome independence as a token of nonlocality.
Thus, when (\ref{eq:Gri24}) is correctly interpreted, it is unjustified to consider it an inconsistent application of classical physics to quantum mechanics.
\section{Quantum Common Causes}\label{sec:QCC}
According to Griffiths, we can explain quantum mechanics locality through the presence of \emph{quantum common causes} (QCC) instead of \emph{classical common causes} (CCC).
He explains the role of QCC:
\begin{quote}
\emph{``Experiments that test Bell inequalities using entangled photon pairs already assume a common cause in the sense that pairs of photons produced at the source in the same, rather than a different, down conversion event are identified using their arrival times. All that is needed in addition is an argument that the polarizations measured later were also created in the same (local) event.''}
\end{quote}
Griffiths's above argument sounds very much like a classical causal explanation of the kind Bell and Einstein would have wanted.
Whence, the role of his QCC would be the same as the CCC represented by $\lambda$ in Bell's formulation.

The difference between CCC and QCC is, according to Griffiths, the violation of statistical independence since he says
\emph{``...where proper use was made of a genuinely quantum ``hidden variable'' $\lambda$, as an example of a ``quantum cause'', in the same sense as that employ here.''}
Then he explains very clearly that his quantum common causes violates the statistical independence hypothesis: \emph{``To summarize, the fundamental difficulty with the factorization condition (24)} [(\ref{eq:Gri24}) in this paper]\emph{ is that it assumes a single sample space of mutually exclusive possibilities, independent of $a$ and $b$, with elements labeled by $\lambda$.''}

He is correct, the ``single sample space'' is equivalent to a joint probability \cite{pFin82}. It means that when Alice's and Bob's settings are $a_i, b_k;\,\,i,k\in\{1,2\}$, we should have
\begin{equation}
Pr(\lambda|a_i, b_k)=\Pr(\lambda)\label{eq:mi}
\end{equation}
(\ref{eq:mi}) is the statistical independence hypothesis related to the freedom assumption in Bell's framework.
Whence, according to Griffiths, common causes that violate statistical independence are quantum common causes instead of classical common causes.

However, there are two issues with Griffiths's view on common causes.
First, given that the inequality rests on the two hypotheses, BLC, and statistical independence, rejecting the last one allows to retain the first, and NSL becomes superfluous.
Thus, here Griffiths switches from a no-signaling position to accepting BLC.

Secondly, the rejection of statistical independence and the admission of additional parameters foreign to quantum theory is widely considered an ontological position that is adverse to orthodox quantum mechanics  \cite{pHoo21}.
So this correct solution of quantum nonlocality contradicts Griffiths's views of maintaining the Hilbert-space quantum mechanics as opposed, and incompatible, with classical physics.
\section{Conclusions}
The lesson to be learned from the Bell inequality regarding the local character of quantum mechanics is that we have two options for understanding quantum baffling correlations without implying ``spooky'' actions at a distance.
Either we reconsider local causality rejecting BLC and accepting NSL, or we accept BLC rejecting statistical independence returning to classical notions of causality.

Relativity has taught us that simultaneity is baffling because we do not have a direct intuition of it.
According to Poincar\'{e}, \emph{``If we think we have this intuition, this is an illusion''} \cite{pPoi14}.
Perhaps quantum mechanics is teaching us a similar lesson with regards to locality.
We do not have a direct intuition of nonlocal influences.
So,  maybe we do not need a return to classical ideas if we reject BLC and accept no-signaling as the correct concept of locality compatible with relativity and quantum mechanics.

However, the issue is not closed.
There exists the other view: the rejection of statistical independence \cite{bHal16,pHos20,pHoo21} with an eventual return to classical physics. \footnote{There is a third position advocated, for instance, by Tim Maudlin \cite{pMau14}, Travis Norsen \cite{pNor07}, and Bohmians, considering that nature is irremediably nonlocal.}

In any case, John Stewart Bell asked relevant questions concerning quantum mechanics interpretation.
The Bell theorem is a meaningful insight more pertinent than the naive tautological statement ``quantum is not classical''.
It helps us probing into quantum mechanics' nature and has implications for quantum technologies.
Thus when correctly interpreted and stripped away of unnecessary metaphysical burden, it does not reduce to a mere inconsistent application of classical physics to Hilbert-space quantum mechanics.

Finally, it is worth highlighting the irony in the widespread belief sustaining that, to retain local causality, it suffices to forgo classical realism.
Quite the contrary, as we have seen, retaining quantum locality (in the BLC sense) requires a regression to classical physics and the acceptance of ``spooky'' hidden variables instead of ``spooky'' action at a distance.

%\begin{acknowledgments}
%\end{acknowledgments}
%\appendix
%\section{}
%\appendix*
%\section{Proof of theorem}
\bibliography{01GC-pra}

\begin{thebibliography}{10}

\bibitem{pGri20}
Robert Griffiths.
\newblock Nonlocality claims are inconsistent with {H}ilbert-space quantum
  mechanics.
\newblock {\em Physical Review A}, 101:022117, 2020.

\bibitem{pBSHC85}
J.S. Bell, A.~Shimony, M.A. Horne, and J.F. Clauser.
\newblock An exchange on local beables.
\newblock {\em Dialectica}, 39(2):85--110, 1985.

\bibitem{pBel64}
J.~S. Bell.
\newblock On the {E}instein-{P}odolsky-{R}osen paradox.
\newblock {\em Physics}, 1:195--200, 1964.

\bibitem{pEPR35}
A.~Einstein, B.~Podolski, and Rosen N.
\newblock Can quantum-mechanical description of physical reality be considered
  complete?
\newblock {\em Phys.Rev.}, 47:777--780, 1935.

\bibitem{pCHSH69}
J.F. Clauser, M.A. Horne, A.~Shimony, and R.A. Holt.
\newblock Proposed experiment to test local hidden-variables theories.
\newblock {\em Phys.Rev.Lett.}, 23:640--657, 1969.

\bibitem{pLam17}
J.~P. Lambare.
\newblock On the {CHSH} form of {B}ell's inequalities.
\newblock {\em Found. Phys.}, 47:321--326, 2017.

\bibitem{pLam21a}
Justo~Pastor Lambare.
\newblock Bell inequalities, counterfactual definiteness and falsifiability.
\newblock {\em International Journal of Quantum Information}, 19(03):2150018,
  2021.

\bibitem{pLam21b}
J.~P. Lambare and R.~Franco.
\newblock A {N}ote on {B}ell's {T}heorem {L}ogical {C}onsistency.
\newblock {\em Found Phys}, 51(84), 2021.

\bibitem{pLam20b}
J.~P. Lambare.
\newblock Comment on ``{A} {L}oophole of {A}ll ``{L}oophole-free''
  {B}ell-{T}ype {T}heorems''.
\newblock {\em Found Sci}, https://doi.org/10.1007/s10699-020-09695-9, 2020.

\bibitem{bShi84b}
A.~Shimony.
\newblock {\em Foundations of Quantum Mechanics in Light of the New
  Technology}, chapter Controllable and uncontrollable non-locality.
\newblock The Physical Society of Japan, 1984.

\bibitem{bHal16}
Michael J.~W. Hall.
\newblock {\em The Significance of Measurement Independence for Bell
  Inequalities and Locality}, pages 189--204.
\newblock Springer International Publishing, Cham, 2016.

\bibitem{pHos20}
Sabine Hossenfelder and Tim Palmer.
\newblock Rethinking superdeterminism.
\newblock {\em Frontiers in Physics}, 8:139, 2020.

\bibitem{pKhr19}
Andrei Khrennikov.
\newblock Get {R}id of {N}onlocality from {Q}uantum {P}hysics.
\newblock {\em Entropy}, 21(8):806, 2019.

\bibitem{pSza94}
L\'{a}szl\'{o}~E. Szab\'{o}.
\newblock On the real meaning of {B}ell's theorem.
\newblock {\em International Journal of Theoretical Physics}, 33:191--197,
  1994.

\bibitem{pKup18}
Marian Kupczynski.
\newblock Closing the door on quantum nonlocality.
\newblock {\em Entropy}, 20(11), 2018.

\bibitem{pWer14}
R.~F. Werner.
\newblock Comment on ``{W}hat {B}ell did''.
\newblock {\em J. Phys. A}, 47:424011, 2014.

\bibitem{pBou17}
S.~Boughn.
\newblock Making sense of {B}ell's theorem and quantum nonlocality.
\newblock {\em Found. of Phys.}, 47:640--657, 2017.

\bibitem{pNor07}
T.~Norsen.
\newblock Against ``realism''.
\newblock {\em Found. Phys.}, 37:311--454, 2007.

\bibitem{pGis12}
Nicolas Gisin.
\newblock {Non-realism: {D}eep {T}hought or a {S}oft {O}ption?}
\newblock {\em Foundations of Physics}, 42:80--85, 01 2012.

\bibitem{pMau14a}
Tim Maudlin.
\newblock Reply to comment on {W}hat {B}ell did.
\newblock {\em J. Phys. A}, 47:424012, 2014.

\bibitem{pLau18}
F.~Laudisa.
\newblock Stop making sense of {B}ell's theorem and nonlocality?
\newblock {\em European Journal for Philosophy of Science}, 8:293--306, 2018.

\bibitem{pCav14}
Eric~G. Cavalcanti and Raymond Lal.
\newblock On modifications of {R}eichenbach's principle of common cause in
  light of {B}ell's theorem.
\newblock {\em Journal of Physics A: Mathematical and Theoretical}, 47:424018,
  2014.

\bibitem{bBac09}
Guido Bacciagaluppi and Antony Valentini.
\newblock {\em Locality and incompleteness}, pages 175--183.
\newblock Cambridge University Press, 2009.

\bibitem{pFra82}
B.~C. Van~Fraassen.
\newblock The {C}harybdis of realism: epistemological implications of {B}ell's
  inequality.
\newblock {\em Synthese}, 52:25--38, 1982.

\bibitem{pFin82}
A.~Fine.
\newblock Hidden variables, joint probability, and the {B}ell inequalities.
\newblock {\em Phys. Rev. Lett.}, 48:291--295, 1982.

\bibitem{pHoo21}
Gerard 't~Hooft.
\newblock Fast {V}acuum {F}luctuations and the {E}mergence of {Q}uantum
  {M}echanics.
\newblock {\em Foundations of Physics}, 51, May 2021.

\bibitem{pPoi14}
Henri Poincar\'{e} and Josiah Royce.
\newblock {\em The Measure of Time}, pages 223--234.
\newblock Cambridge Library Collection - History of Science. Cambridge
  University Press, 2014.

\bibitem{pMau14}
Tim Maudlin.
\newblock What {B}ell did.
\newblock {\em J. Phys. A}, 47:424010, 2014.

\end{thebibliography}
\end{document}